\documentclass[%
preprint,
 amsmath,amssymb,
aps,
]{revtex4-1}

\usepackage{bm}
\usepackage{graphicx}
\usepackage{color}

\begin{document}

\title{Second-phase nucleation on an edge dislocation}

\author{A. R. Massih}
 \altaffiliation[Also at ]{Quantum Technologies AB, Uppsala Science Park, SE-751 83 Uppsala, Sweden}
\affiliation{%
Malm\"{o} University, SE-205 06 Malm\"{o}, Sweden\\
}%




\date{\today}

\begin{abstract}
 A model for nucleation of second phase at or around dislocation in a crystalline solid is considered. The model employs the Ginzburg-Landau theory of  phase transition comprising the sextic term in order parameter ($\eta^6$) in the Landau free energy. The ground state solution of the linearized time-independent Ginzburg-Landau equation has been derived, through which the spatial variation of the order parameter has been delineated. Moreover, a generic phase diagram indicating a tricritical behavior near and away from the dislocation is depicted. The relation between the classical nucleation theory and the Ginzburg-Landau approach has been discussed, for which the critical formation energy of nucleus is related to the maximal of the Landau potential energy. A numerical example illustrating the application of the model to the case of nucleation of hydrides in zirconium alloys is provided.

\end{abstract}

\maketitle

\section{Introduction}
\label{intro}
Nucleation of second phase in the vicinity of elastic defects such as dislocations occurs in many alloys \cite{Boulbitch_Toledano_1998,Leonard_Desai_1998}. For example, in Al-Zn-Mg alloys, dislocations not only induce and enhance nucleation and growth of the coherent Laves phase MgZn$_2$ precipitates, but also produce a spatial precipitate size gradient around them   \cite{Allen_Vandesande_1978,Deschamps_et_al_1999,Deschamps_Brechet_1999}. Another example is formation of a new phase in ammonium bromide (NH$_4$Br), namely $\beta\to\gamma$ phase transition, which is observed to occur in the vicinity of crystal dislocations \cite{Belousov_Volf_1980}. In titanium and zirconium alloys, used in aerospace and nuclear industries, the presence of hydrogen leads to hydride formation (TiH$_x$, ZrH$_x$) close to and on dislocations, causing embrittlement of the alloy, thereby reducing its performance during service \cite{Chen_Li_Lu_2004,Perovic_Weatherly_1984}.

Cahn \cite{Cahn_1957} provided the first quantitative model for nucleation of second phase on dislocations in solids using classical nucleation theory.  Cahn's model assumes that a cross-section of the nucleus is circular, which is valid for a screw dislocation. Moreover, it posits that the nucleus is incoherent with the matrix. The issue of the formation of coherent nucleus on or near an edge dislocation has been studied theoretically by Lyubov \& Solovyev \cite{Lyubov_Solovyev_1965} and Dollins \cite{Dollins_1970}. These theoretical approaches have been thoroughly appraised by Larch\'e \cite{Larche_1979}. In a recent study, Hin and coworkers \cite{Hin_et_al_2008a} studied heterogeneous precipitation of FeC particles on dislocations in the iron-carbon binary system using kinetic Monte Carlo technique.

Here, we  present a generic model for nucleation of a new phase near edge dislocations in crystals. The model rests on the Ginzburg-Landau theory of phase transition in which an order parameter designates the symmetry of the system. Moreover, the elastic property of the solid is taken into consideration by the \emph{striction} term in the free energy, which accounts for the interaction between the order parameter and deformation \cite{Larkin_Pitkin_1969,Imry_1974}. The model is in line with earlier approaches by Nabutovskii \& Shapiro \cite{Nabutovskii_Shapiro_1977} and Boulbitch \& Toledano \cite{Boulbitch_Toledano_1998}; where herein note the case of phase transition near edge dislocations has been elaborated. The model is pertinent to systems in which second phase nucleation is accompanied by a preferred orientation of nuclei under external force. This includes $\alpha^{\prime\prime}$-phase precipitation in Fe-N alloys \cite{Sauthoff_1981}, $\theta^\prime$-phase nucleation in Al-Cu alloys \cite{Li_Chen_1998}, $\delta$-hydride formation in Zr-alloys containing hydrogen \cite{Massih_Jernkvist_2009}. We should, however, point out that in this note, we only treat the details of the ordering (orientation) aspect of the problem, which is characterized by a non-conserved order parameter. That is, the effect of composition field is decoupled from the Ginzburg-Landau model. A more general formulation with coupled non-conserved order parameter, conserved variable (concentration) and elastic field was presented elsewhere \cite{Massih_2011a}, see also \cite{Li_Chen_1998}.

The paper is organized as follows. The model set up and the basic equations are described in section \ref{sec:gov-eqs}. The ground state solution of the linearized steady-state Ginzburg-Landau equation, in the vicinity of edge dislocation, is derived in section \ref{sec:bstate} using the method of Dubrovskii \cite{Dubrovskii_1997}. The phase diagram ensued from the model is presented in section \ref{sec:phase-equilib}. Section \ref{sec:cnt} discusses the relation between the present model and the classical nucleation theory. That section also includes a numerical example pertinent to nucleation of hydrides in zirconium alloys. Finally, in section \ref{sec:close}, we end the paper with some concluding remarks. Some mathematical details are presented in the appendices.

\section{Model description}
\label{sec:gov-eqs}

\begin{figure}[htbp]
\begin{center}
\includegraphics[width=0.70\textwidth]{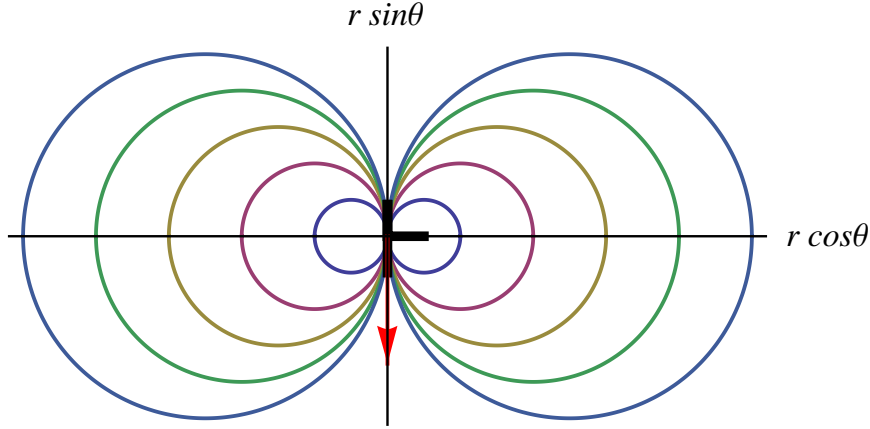}\\
\caption{Equipotentials of an edge dislocation with interaction energy given by  Eq. (\ref{eqn:disl-potential}). The downward arrow indicates the Burgers vector.}
\label{fig:equipot}
\end{center}
\end{figure}

We consider an edge dislocation, the line of which coincides with the 0z axis and the Burgers vector with components $b_y=-b,b_x=b_z=0$, see Fig \ref{fig:equipot}. In an elastically isotropic crystal, such a dislocation creates a deformation potential \cite{Slyusarevl_Chishko_1984}
\begin{equation}
V(r,\theta) = \mathcal{B}\frac{\cos\theta}{r},
\label{eqn:disl-potential}
\end{equation}
\noindent
where $(r,\theta)$ are the polar coordinates at the point of observation in a plane perpendicular to the dislocation line. Here $\mathcal{B}$ is a material dependent parameter denoting the strength of the potential. It is related to the basic constants of the metal, namely, $\mathcal{B}=b\varepsilon_F(1-2\nu)/[3\pi(1-\nu)]$, where $\varepsilon_F$ is the Fermi energy and $\nu$ is Poisson's ratio.

The considered phase transformation is the nucleation of second phase in a solid solution. It is characterized by an order parameter accounting for the symmetry of the structure (system) under consideration. In general, it is defined by a vector field $\boldsymbol{\eta}(\mathbf{r},t)$ being a function of space $\mathbf{r}$ and time $t$. We write the total free energy for the system
\begin{eqnarray}
\mathcal{F} = \mathcal{F}_{st}+\mathcal{F}_{el}+\mathcal{F}_{int},
\label{eqn:toten}
\end{eqnarray}
\noindent
where $\mathcal{F}_{st}$ is the structural free energy, $\mathcal{F}_{el}$ the elastic strain energy, and  $\mathcal{F}_{int}$ is the interaction energy between the structural order parameter and the strain field. The structural free energy is
\begin{eqnarray}
\mathcal{F}_{st}= \int
\big[\frac{g}{2}(\nabla\eta)^2+\mathcal{V}(\eta)\big]d\mathbf{r}.
\label{eqn:f-str}
\end{eqnarray}
where the space integral is within the volume of the system. Here  $g(\nabla\eta)^2$ accounts for the spatial dependence of the order parameter, $g$ is a positive constant, and  $\mathcal{V}(\eta) = \frac{1}{2}r_0\eta^2 + \frac{1}{4}u_0\eta^4 + \frac{1}{6}v_0\eta^6$ is the Landau type potential energy, where $r_0$ and $u_0$ are, in general, functions of temperature and external field, here stress. The coefficient $v_0$ is taken to be a positive constant. Its role is to ensure stability.  Also, we wrote $\eta^2=\boldsymbol{\eta}\cdot\boldsymbol{\eta}$ and so on.  The elastic free energy is
\begin{eqnarray}
\mathcal{F}_{el}= \int \Big[\frac{K}{2}\big(\nabla \cdot \mathbf{u}\big)^2
+ M\sum_{ij}\Big(u_{ij} - \frac{\delta_{ij}}{d}\nabla \cdot \mathbf{u}\Big)^2\Big]d\mathbf{r},
\label{eqn:f-el}
\end{eqnarray}
\noindent
where $K$ and $M$ are the bulk and shear modulus, respectively, $u_{ij}=(\nabla_j u_i+\nabla_i u_j)/2$ is the strain tensor with $\nabla_i\equiv \partial/\partial x_i$, $d$ the space dimensionality, and $i,j$ stand for $x,y,z$ in $d=3$ ($x,y$ in $d=2$). Finally, the interaction energy is
\begin{equation}
\mathcal{F}_{int} =\kappa\int  \eta^2\, \nabla \cdot \mathbf{u}\, d\mathbf{r},
\label{eqn:f-int}
\end{equation}
where  $\eta^2\nabla \cdot \mathbf{u}$ accounts for the interaction between deformation and the order parameter and the constant $\kappa$, referred to as the striction factor, denotes the strength of this interaction. For structural orientation of the second phase, the order parameter may supposed to be a two-component vector field $\boldsymbol{\eta}=(\eta_1,\eta_2)$, where $(\eta_1,\eta_2)=(\pm\eta_0,0)$ would denote one preferred orientation of nuclei (precipitates) and $(\eta_1,\eta_2)=(0,\pm\eta_0)$ another, where $\eta_0$ is a non-zero constant \cite{Li_Chen_1998}. Furthermore, $\boldsymbol{\eta}=(0,0)$ would describe the solid solution with no precipitates present. Here, for the sake of simplicity, we assume that the order parameter is a scalar field (an Ising model), taking values of $\eta=0$ (solid solution) or $\eta \ne 0$ (nucleus)

The evolution of the order parameter is described by the time-dependent Ginzburg-Landau (GL) type equation
\begin{equation}
\label{eqn:tdgl-eq}
\frac{\partial \eta}{\partial t}  =  -L_a\frac{\delta\mathcal{F}}{\delta \eta},
\end{equation}
where $L_a$ is the mobility; also the effect of the background random thermal (Langevin) noise is ignored. This is the basic kinetic equation for a non-conserved field, and it corresponds to the model A in the classification of Hohenberg \& Halperin \cite{Hohenberg_Halperin_1977}. Moreover, defects in a crystal such as dislocations give rise to internal strains in the solid affecting the mechanical equilibrium condition. The equilibrium condition in the presence of an edge dislocation and the force field generated by the order parameter is expressed as (see Appendix \ref{sec:appA} and \cite{Landau_Lifshitz_1970})
\begin{eqnarray}
M\nabla^2 \mathbf{u}+(\Lambda-M)\nabla \nabla \cdot \mathbf{u} + \kappa\nabla \eta^2 = -Mb\boldsymbol{e}_y\delta(x)\delta(y),
\label{eqn:mecheq}
\end{eqnarray}
where $\Lambda \equiv K+2M(1-1/d)$,  $b$ is the magnitude of the Burgers vector, $\boldsymbol{e}_y$ denotes the unit vector along the $y$ axis, and $\delta(\bullet)$ is the Dirac delta. Equation (\ref{eqn:mecheq}) is used to solve $\mathbf{u}$ in terms of $\eta^2$ and the displacements of the dislocation; thereafter, we eliminate the elastic field $\mathbf{u}$  from the expression for the Euler-Lagrange condition on the free energy (Appendix \ref{sec:appA}). Hence, the total energy can be expressed as \cite{Massih_2011a}
\begin{equation}
\mathcal{F}[\eta] = \mathcal{F}_0+\int \Big[\frac{g}{2}(\nabla\eta)^2+\frac{1}{2}r_1\eta^2 + \frac{1}{4}u_1\eta^4 + \frac{1}{6}v_0\eta^6\,\Big] d\mathbf{r},
\label{eqn:toten1}
\end{equation}
where $\mathcal{F}_0=\mathcal{F}[0]$ is a function of temperature and stress, $r_1=r_0-B\cos\theta/|\mathbf{r}|$, $r_0=\alpha(T/T_c-1)\equiv \alpha\tau$, $\alpha$ is a positive constant, $T$ the temperature,  $T_c$ the phase transition temperature in the absence of elastic coupling, $u_1=u_0-2\kappa^2/\Lambda$ for an edge dislocation embedded in the matrix, $u_0$ can be both positive and negative, and $d\mathbf{r}=rdrd\theta dz$. Also, $B=\kappa A$ and $A=(2b/\pi)M/\Lambda\equiv (b/\pi)(1-2\nu)/(1-\nu)$. We regard $\mathcal{F}[\eta]$ to have dimensions of energy $[\mathrm{M}\mathrm{L}^{2}\mathrm{T}^{-2}]$. By considering $\eta$ to have dimensions of inverse area $[\mathrm{L}^{-2}]$, the variable $r_1$ gets dimensions $[\mathrm{M}\mathrm{L}^{3}\mathrm{T}^{-2}]$, and so on. Some of the constants appearing in the coefficients of the powers of $\eta$, such as $T_c$, the elasticity constants, the Burgers vector, are directly measurable for a particular system. Other parameters, such as $g$, $\alpha_0$, $v_0$, $\kappa$, if not directly measurable, may be determined from \emph{ab initio} type methods; see also section \ref{sec:discuss}.

Let us first express  Eq. (\ref{eqn:toten1}) in dimensionless form
\begin{equation}
\mathcal{F}[\psi] =F_0\int \Big[(\nabla_\rho\psi)^2 + \frac{1}{2}\big(U(\boldsymbol{\rho})-E\big)\psi^2 + \frac{1}{2}\psi^4 + \frac{1}{6}\psi^6\,\Big] d\boldsymbol{\rho},
\label{eqn:toten1-dimless}
\end{equation}
with $d\boldsymbol{\rho}= \rho d\rho d\theta d\zeta$ and the introduced dimensionless parameters
\begin{equation}
\rho = r/\ell, \quad \psi = -\eta\sqrt{u_1g}/2B, \quad \zeta = z/\ell, \quad U(\boldsymbol{\rho})=-\cos{\theta}/\rho,
\label{eqn:paradimless1}
\end{equation}
\noindent
\begin{equation}
E = -r_0g/2B^2, \quad \ell = g/2B, \quad F_0 = Bg/u_1, \quad v_0 \equiv gu_1^2/8B^2.
\label{eqn:paradimless2}
\end{equation}

Here, we treat nucleation in the presence of the edge dislocation in thermal-mechanical equilibrium.  Furthermore, we suppose that there is no Langevin noise in the system. Hence Eq. (\ref{eqn:tdgl-eq}), using Eq. (\ref{eqn:toten1-dimless}), reduces to
\begin{eqnarray}
\label{eqn:gl-eq}
 \nabla^2\psi & = & \big(U(\rho,\theta)-E\big)\psi+2\psi^3+\psi^5,\\
 \text{where} \quad \nabla^2 & = & \frac{\partial^2 }{\partial \rho^2}+\frac{1}{\rho}\frac{\partial}{\partial \rho}+\frac{1}{\rho^2}\frac{\partial^2}{\partial\theta^2},
 \label{eqn:laplace-2d}
\end{eqnarray}
and the boundary conditions: $\psi(\rho=\infty,\theta)=0$ and $\psi(\rho=0,\theta)\ne \infty$.

One can show that for a certain $E<0$, i.e. $T>T_c$, Eq. (\ref{eqn:gl-eq}) has a nontrivial solution for $E>E_0$ ($T<T_0$). In the next section, we shall specify and calculate $E_0$ and obtain the bound state solutions of Eq. (\ref{eqn:gl-eq}) as $E \to E_0$.

\section{Bound states}
\label{sec:bstate}

The value of $E=E_0$ is found from Eq. (\ref{eqn:gl-eq}) for which a solution would first emerge. This value is determined from the solution of the linearized  Ginzburg-Landau equation, viz.
\begin{equation}
\label{eqn:lgl-eq}
 \nabla^2\psi  =  \big(U(\rho,\theta)-E\big)\psi,
\end{equation}
\noindent
which is the lowest ``level'' of Eq. (\ref{eqn:gl-eq}), and it corresponds to the Gaussian approximation of the Ginzburg-Landau free energy functional. Equation (\ref{eqn:lgl-eq}) is equivalent to the Schr\"odinger equation in two dimensions.  The bound state solution $\psi_E$ to this equation exists if $\mathcal{F}[\psi_E]<0$
 and that occurs at $E=E_0$ \cite{Nabutovskii_Shapiro_1978}.

The variables in Eq. (\ref{eqn:lgl-eq}) with $U(\rho,\theta)$ defined in Eq. (\ref{eqn:paradimless1}) do not separate. To solve Eq. (\ref{eqn:lgl-eq}) we chose the method proposed by Dubrovskii \cite{Dubrovskii_1997} outlined in Appendix \ref{sec:appB}. Recall that the eigenfunctions of the equation must satisfy the aforementioned boundary condition on $\psi$. So we may write Eq. (\ref{eqn:2dse-cef}) in Appendix \ref{sec:appB}, expressed in terms of  cosine-elliptic Mathieu function $ce_m(a,\vartheta,q)$, cf. Appendix \ref{sec:appC}, as
\begin{equation}
\label{eqn:2dse-cefe}
 \psi_{nm}  =  \frac{A_{nm}}{\sqrt{\pi}} ce_m(a,\vartheta,q)\rho^{\mu} \exp(-\beta_n\rho){_1\!F_1}(-n,2\mu+1,2\beta_n \rho),
\end{equation}
\noindent
where $A_{nm}$ is the normalization constant, $\mu =\sqrt{a+\alpha_1}/2$, $\vartheta=(\theta-\pi)/2$,  $a$ is a separation constant, $q$, $\alpha_1$ are variational constants to be determined, $\beta_n$ is a certain function of $q$, $n$ is an integer, $m$ is an even integer, and ${_1\!F_1}(\bullet,\bullet,\bullet)$ is the confluent hypergeometric function of the first kind. We should point out that for each $q$, the separation constant $a=a_m$ assumes an infinite set of discrete values depending on the parity and the index of the function $ce_m(a,\vartheta,q)$, cf. Appendix \ref{sec:appC}.

\begin{figure}[htbp]
\begin{center}
\includegraphics[width=0.70\textwidth]{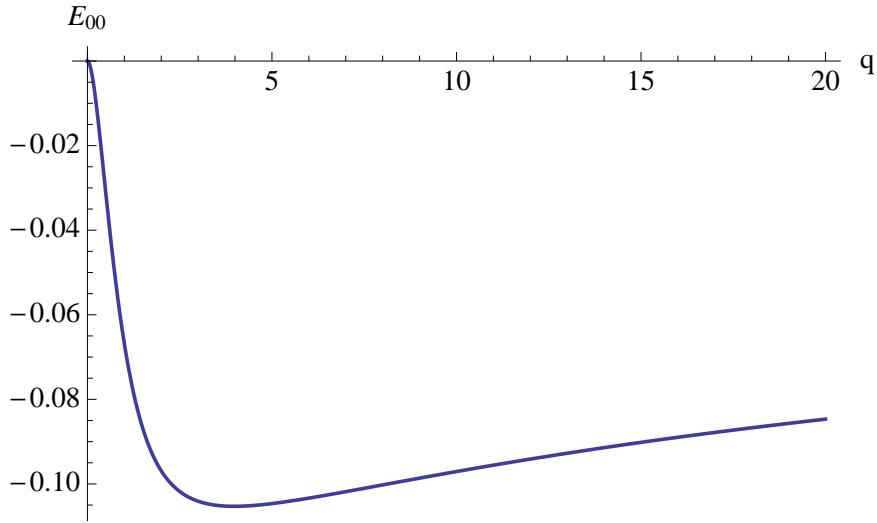}\\
\caption{The ground state energy of the total Hamiltonian, Eq. (\ref{eqn:2dse-tot-ev2}), versus $q$. The minimum occurs at $q_{\mathrm{min}}=3.918$ and $E_{00}(q_{\mathrm{min}})=-0.105$, cf. Table \ref{minimal-par}.}
\label{fig:E00-plot}
\end{center}
\end{figure}

\begin{figure}[htbp]
\begin{center}
\includegraphics[width=0.80\textwidth]{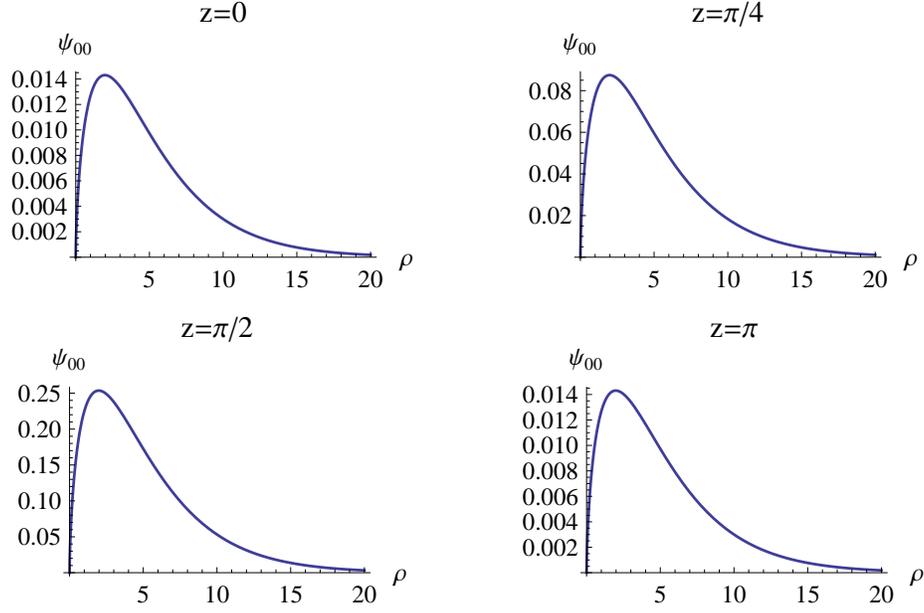}\\
\caption{The ground state eigenfunction $\psi_{00}(\rho,\mathrm{z},q_{\mathrm{min}})$ with $q_{\mathrm{min}}=3.918$, cf. Eq. (\ref{eqn:2dse-cef-gs}) and Table \ref{minimal-par}.}
\label{fig:psi00-plot}
\end{center}
\end{figure}

\begin{figure}[htbp]
\begin{center}
\includegraphics[width=0.65\textwidth]{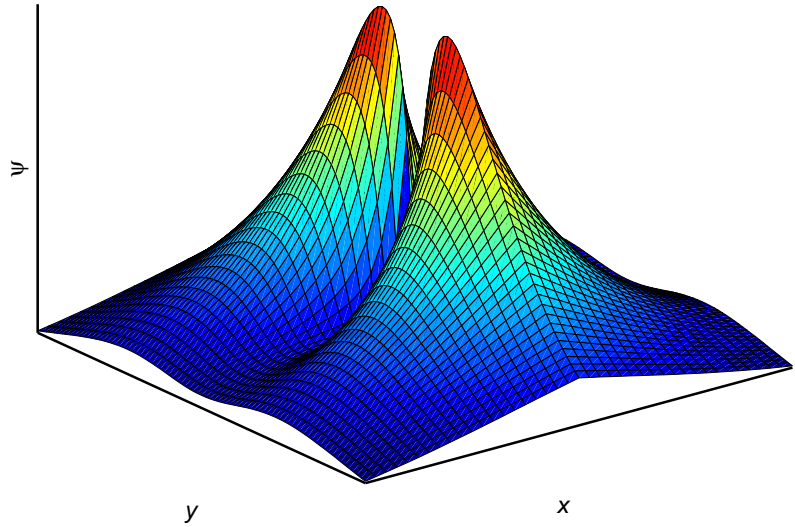}\\
\caption{The distribution of the ground state eigenfunction $\boldsymbol{\psi} \equiv \psi_{00}(\rho,\mathrm{\vartheta},q_{\mathrm{min}})$ with $q_{\mathrm{min}}=3.918$, cf. Eq. (\ref{eqn:2dse-cef-gs}) and Table \ref{minimal-par}, corresponding to the spatial variation of order parameter near the edge dislocation. Here $x=\rho\cos\vartheta$ and $y=\rho\sin\vartheta$.}
\label{fig:psi-xy}
\end{center}
\end{figure}

The composite ground state eigenfunction corresponding to Eq. (\ref{eqn:2dse-cefe}) is then
\begin{equation}
\label{eqn:2dse-cef-gs}
 \psi_{00}(\rho,\vartheta,q)  = \frac{A_{00}}{\sqrt{\pi}}ce_0(a_0,\vartheta,q)\rho^{\mu} \exp(-\beta\rho),
\end{equation}
\noindent
 where we put ${_1\!F_1}(0,2\mu+1,2\beta \rho)=1$ and $\beta \equiv \beta_0$. We should note that $\beta$ and $\mu$ are functions of $q$, $\alpha_1$ and $\alpha_2$ (see below). The normalization constant is evaluated to be
\begin{equation}
\label{eqn:2dse-a00}
 A_{00}  =  \sqrt{\frac{2(2\beta)^{2+2\mu}}{\Gamma(2+2\mu)}}.
 \end{equation}
 \noindent
The ground state eigenvalue of the total Hamiltonian operator $\hat{H}=\nabla^2 -U(\rho,\vartheta)$, to be minimized, is expressed as ($E_{00}
\Leftrightarrow E_0$)
\begin{equation}
\label{eqn:2dse-tot-ev}
 E_{00}  \leqq  \langle 00|\hat{H}|00\rangle \equiv \int_0^\infty \rho d\rho \int_{-\pi/2}^{\pi/2} \psi_{00}(\rho,\vartheta) \hat{H}\psi_{00}(\rho,\vartheta)d\vartheta.
 \end{equation}
 \noindent
Next, the integration of the right hand side of Eq. (\ref{eqn:2dse-tot-ev}) over the variable $\vartheta$ yields
\begin{eqnarray}
\label{eqn:2dse-tot-ev2}
E_{00}  &\leqq& \frac{(2\beta)^{2+2\mu}}{\Gamma(2+2\mu)} \int_0^\infty \rho^{\mu+1} \exp(-\beta\rho) \hat{\mathcal{K}}\rho^{\mu}\exp(-\beta\rho)d\rho,\\
 \mbox{with} \quad \hat{\mathcal{K}} &=&  - \frac{d^2 }{d \rho^2}  - \frac{1}{\rho}\frac{d }{d \rho} +\frac{K_0}{4\rho^2}-\frac{K_0^2-a_0(q)}{2q\rho},
 \end{eqnarray}
 \noindent
 where $a_0(q)<0$ is the eigenvalue of Eq. (\ref{eqn:mathieu}) in Appendix \ref{sec:appB}, corresponding to the eigenfunction $ce_0(a,\vartheta,q)$ and
 \begin{equation}
\label{eqn:2dse-K0}
 K_0^2  =  -\frac{2}{\pi}\int_{-\pi/2}^{\pi/2} ce_0(a_0,\vartheta,q) \frac{d^2 }{d \vartheta^2} ce_0(a_0,\vartheta,q) d\vartheta.
 \end{equation}
 \noindent
Now, $\hat{\mathcal{K}}\phi(\rho)=E_{00}\phi(\rho)$ with $\phi(\rho)= \rho^{\mu} \exp(-\beta\rho)$ leads to
\begin{equation}
\label{eqn:2dse-param}
 \beta  =  \frac{K_0^2-a_0(q)}{2q(K_0+1)}, \quad \mu =\frac{K_0}{2}, \quad E_{00}=-\beta^2.
 \end{equation}
 \noindent
Additional analysis \cite{Dubrovskii_1997} yields (cf. Appendix \ref{sec:appB})
\begin{equation}
\label{eqn:2dse-gen-par1-gs}
 \alpha_1 = K_0^2-a_0, \quad \alpha_2 = \frac{\alpha_1}{2q}.
\end{equation}
where  $\alpha_2$  is an additional variational constant related to $\alpha_1$. Figure \ref{fig:E00-plot} depicts $E_{00}(q)$ as a function of $q$. The obtained numerical values for the parameters, upon minimization of $E_{00}(q)$, are listed in Table \ref{minimal-par}. And in Figure \ref{fig:psi00-plot}, we  depict the radial dependence of $\psi_{00}(\rho,\vartheta)$ at $q=q_{\mathrm{min}}$. The eigenfunction $\psi_{00}(\rho,\vartheta)$, corresponding to the normalized order parameter in the vicinity of the dislocation, is shown in Figure \ref{fig:psi-xy} on the $xy$-plane.

\begin{table}[b]
\caption{\label{minimal-par}Numerical values for the parameters obtained by minimization of $E_{00}(q)$, Eq. (\ref{eqn:2dse-tot-ev}), where $q_m=q_{\mathrm{min}}$.}
\begin{ruledtabular}
\begin{tabular}{ccc}
\hline
$q_m=3.918$ & $E_{00}(q_m)=-0.105$ & $K_0(q_m)=1.284$ \\
$A_{00}(q_m)=0.428$ & $\beta(q_m)=0.3245$ & $a_{0}(q_m)=-4.159$\\
$\mu(q_m)=0.642$ 	& $\alpha_1(q_m)=5.808$  &  $\alpha_2(q_m)=0.741$ \\

\end{tabular}
\end{ruledtabular}
\end{table}

\section{Phase equilibria}
\label{sec:phase-equilib}
 Let us study the phase diagram for the system under consideration. To first approximation, as in \cite{Boulbitch_Toledano_1998}, the equilibrium field parameter $\psi$ is taken to be the ground state solution of linearized Eq. (\ref{eqn:gl-eq}), i.e., $\psi  = \psi_{00}(\rho,\vartheta,q_m)$. We then express the order parameter for the system in the form
\begin{equation}
\label{eqn:equilib-sol}
\eta(\rho,\vartheta) =  \mathfrak{m}\frac{A_{00}^\prime}{\sqrt{\pi}}ce_0(a_0,\vartheta,q)\Big(\frac{\rho}{\rho_0}\Big)^{\mu} \exp(-\rho/\rho_0)+\mathcal{O}(\mathfrak{m}^2),
\end{equation}
\noindent
where we used Eq. (\ref{eqn:2dse-cef-gs}), $\mathfrak{m}$ is an amplitude, $\rho_0 \equiv 1/\beta$ and $A_{00}^\prime\equiv A_{00}\rho_0^\mu$.   Note that if $\mathfrak{m}$ is considered to be dimensionless, then the right-hand side of Eq. (\ref{eqn:equilib-sol}) needs to scaled by the factor $\sqrt{u_1g}/2B$ to make $\eta$ to have dimensions of inverse area $\left[\mathrm{L}^{-2}\right]$.

Now substituting Eq. (\ref{eqn:equilib-sol}) into (\ref{eqn:toten1}), using the data in Table \ref{minimal-par}, and integrating over the volume ($0 \le \rho \le \infty$, $0 \le \vartheta \le \pi,0 \le z \le L$) yield the equilibrium free energy of an ordered nucleus around the dislocation. So the free energy is expressed in the form
\begin{eqnarray}
\mathcal{F}(\mathfrak{m}) = \mathcal{F}_0+\ell^2L\Big[\frac{1}{2}(r_0-r_0^\ast)\mathfrak{m}^2+\frac{1}{4}(u_0-u_0^\ast)\langle \psi^4\rangle\mathfrak{m}^4+\frac{1}{6}v_0\langle \psi^6\rangle\mathfrak{m}^6\Big],
\label{eqn:landau-fe}
\end{eqnarray}
where $\mathcal{F}_0$ is the free energy of the defect in the parent phase ($\mathfrak{m} = 0$), $L$ is the size of the crystal in the $z$-direction, $u_0^\ast=2\kappa^2/\Lambda$, and
\begin{eqnarray}
\label{eqn:rostar}
r_0^\ast &=& \Big[\frac{2B}{\ell}c_3-\frac{g}{\ell^2}\mu\big(1+\frac{c_2}{2\mu\pi}\big)\Big]\frac{\rho_0^{-2}}{\mu(1+2\mu)},\\
\langle \psi^{2n}\rangle &=& \int_0^\infty \int_0^\pi\Big[\frac{A_{00}^\prime}{\sqrt{\pi}}ce_0(a_0,\vartheta,q)\Big(\frac{\rho}{\rho_0}\Big)^{\mu} \exp(-\rho/\rho_0)\Big]^{2n} \rho d\rho d\vartheta,
\label{eqn:moment-n}
\end{eqnarray}
where $c_2=2.5903$ and $c_3=1.1642$ are obtained by appropriate integrations carried over the angle ($0\le \vartheta \le \pi$). Also the integrations in Eq. (\ref{eqn:moment-n}) can readily be evaluated, viz.
\begin{eqnarray}
\label{eqn:moment-2}
\langle \psi^{2}\rangle &=& 1,\\
\langle \psi^{4}\rangle &=& 0.1842\frac{4^{1-2\mu}}{\rho_0^2}\frac{\Gamma(2+4\mu)}{\Gamma(2+2\mu)^2},\\
\label{eqn:moment-4}
\langle \psi^{6}\rangle &=& 0.0791\frac{2^7\times 3^{-2-6\mu}}{\rho_0^4}\frac{\Gamma(2+6\mu)}{\Gamma(2+2\mu)^3}.
\label{eqn:moment-6}
\end{eqnarray}

Equation (\ref{eqn:landau-fe}) is a kind of a mean field variant of the Landau free energy \cite{Landau_Lifshitz_1980} with the order parameter $\mathfrak{m}$. The parameter $v_0$ in Eq. (\ref{eqn:landau-fe}) is taken to be a positive fixed constant, but $\bar{r}\equiv (r_0-r_0^\ast)$ and $\bar{u}\equiv (u_0-u_0^\ast)$ are varying parameters, assuming both positive and negative values. The system has a phase transition at $\bar{r}=0$, which can be either of first order or second order depending on sign of $\bar{u}$. If $\bar{u}>0$, then $r_0>r_0^\ast$ gives $\mathfrak{m}=0$, i.e. a situation with no nucleus. At  $r_0=r_0^\ast$, nucleation occurs by second order transition; and for $0<r_0<r_0^\ast$, $\mathfrak{m}\ne0$, implying that the nuclei grow continuously. On the other hand, if $\bar{u}<0$, the transition is first order and the nuclei form on the coexistence line of the phase diagram, see below.

\begin{figure}[htbp]
\begin{center}
\includegraphics[width=0.55\textwidth]{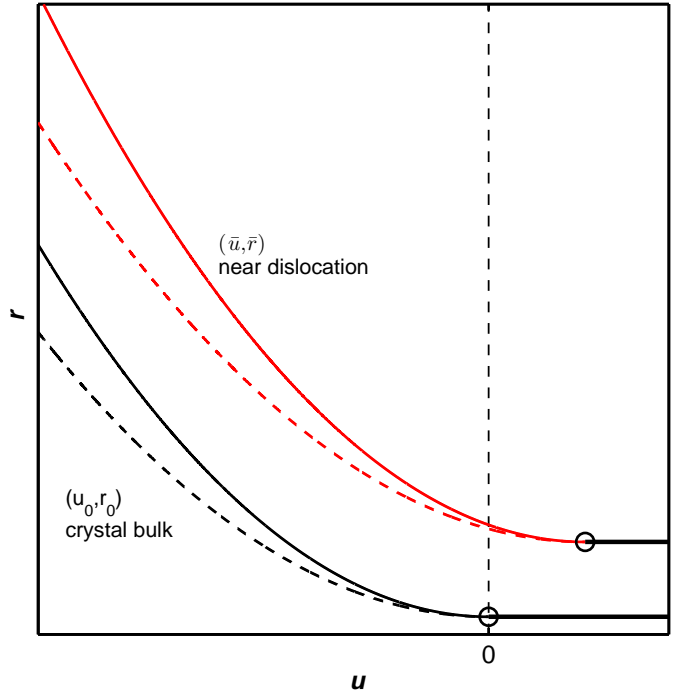}\\
\caption{Phase diagram of nucleation in the bulk of rigid  crystal $(u_0,r_0)$ and near the edge dislocation $(\bar{u},\bar{r})$ obtained from Landau's potential (see figure \ref{fig:Landau-pot}). The dashed line describes the first order transition, the solid line the border of instability, and the open circles are the tricritical points.}
\label{fig:eta6-pd}
\end{center}
\end{figure}

\begin{figure}[htbp]
\begin{center}
\includegraphics[width=0.705\textwidth]{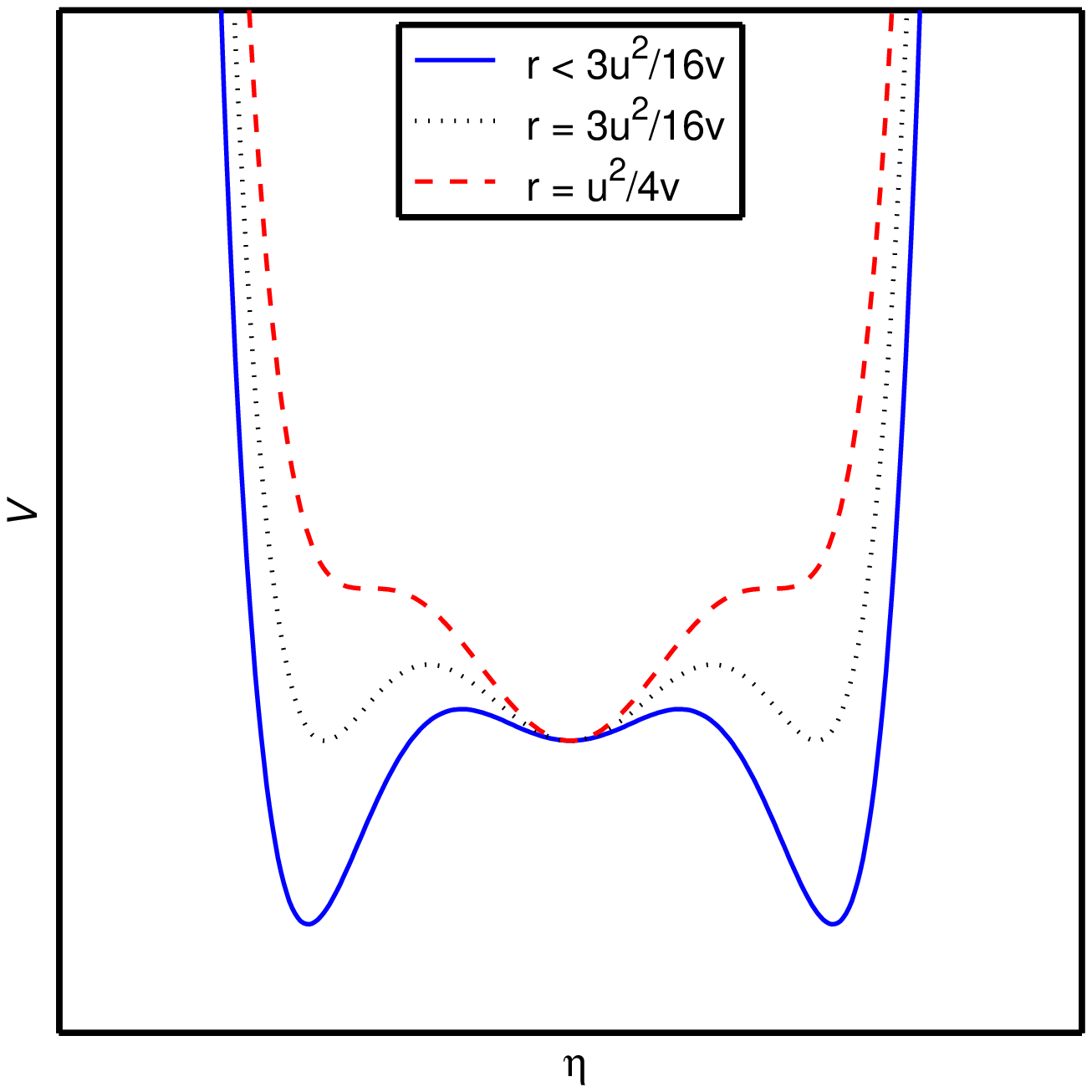}\\
\caption{The Landau potential energy $\mathcal{V} = \frac{r}{2}\eta^2 + \frac{u}{4}\eta^4+\frac{v}{6}\eta^6$, with $v>0$, $u<0$, $r>0$, and $\eta$ scalar. The curves determine the lines plotted in figure \ref{fig:eta6-pd}. The dashed curve marks the emergence of metastable second-phase II (two local minima) in the parent phase I (global minimum); the dotted line indicates that phases I and II are equally stable; and the solid line refers to when phase I is less stable than phase II, which remains in that state until $r \le 0$.}
\label{fig:Landau-pot}
\end{center}
\end{figure}

 The value of $\mathfrak{m}$ can be determined by the equilibrium condition $\mathcal{F}(\mathfrak{m}) = \mathcal{F}_0$, and at the same time  $d\mathcal{F}/d\mathfrak{m} = 0$, which give
\begin{eqnarray}
\label{eqn:op-amp}
\mathfrak{m}^2 = -\frac{3}{4}\frac{\langle \psi^4\rangle}{\langle \psi^6\rangle}\Big(\frac{u_0-u_0^\ast}{v_0}\Big).
\end{eqnarray}
Substituting this in $\mathcal{F}(\mathfrak{m}) = \mathcal{F}_0$, we find
\begin{eqnarray}
\label{eqn:1st-order-line}
r_0 &=& r_0^\ast+\frac{3}{16}\frac{\langle \psi^4\rangle^2}{\langle \psi^6\rangle}\frac{1}{v_0}(u_0-u_0^\ast)^2,\\
\text{with} \quad  r_0^\ast &=& \frac{2}{1+2\mu} \Big[\frac{c_3}{\mu}-\big(1+\frac{c_2}{2\pi\mu}\big)\Big]r_{00}.
\label{eqn:rostar2}
\end{eqnarray}
\noindent
This is the equation for first order phase transition line with $r_{00}$ denoting the ground state value of $r_0$. Next minimizing $\mathcal{F}(\mathfrak{m})$, i.e. $d\mathcal{F}/d\mathfrak{m} = 0$ and $d^2\mathcal{F}/d\mathfrak{m}^2 \ge 0$, then for $r_0<r_0^\ast$ and $u_0<u_0^\ast$, we can calculate the resulting $\mathfrak{m}$ (with admissible solutions); thereby leading to
\begin{eqnarray}
\label{eqn:meta-line}
r_0\leqq r_0^\ast+\frac{1}{4v_0}\frac{\langle \psi^4\rangle^2}{\langle \psi^6\rangle}(u_0-u_0^\ast)^2,
\end{eqnarray}
where the equality marks the line of instability, i.e. the onset of nucleation. Note that for $r_0<r_0^\ast$ and $u_0<u_0^\ast$,  one solution ($\mathfrak{m}_+$) gives two minima while another ($\mathfrak{m}_-$) gives the maximum of the free energy. Also, $(r_0-r_0^\ast)=\mathrm{const.}$ and $u_0>u_0^\ast$ (and $v_0>0$) mark out the second order phase transition line. To illustrate this, we rewrite Eqs. (\ref{eqn:1st-order-line}) and (\ref{eqn:meta-line}) in a more concise form
\begin{eqnarray}
\label{eqn:translines}
\bar{r}=\frac{3}{16v_0}\frac{\langle \psi^4\rangle^2}{\langle \psi^6\rangle}\bar{u}^2,\qquad \bar{r}\leqq \frac{1}{4v_0}\frac{\langle \psi^4\rangle^2}{\langle \psi^6\rangle}\bar{u}^2.
\end{eqnarray}
\noindent
Recalling that in the crystal bulk $r_0=\alpha(T/T_c-1)$, Eq. (\ref{eqn:translines}) together with the line $\bar{r}=0$ (or  $\bar{r}=\mathrm{const.}$) and $\bar{u}>0$, define the phase diagram for the system in the $(\bar{u},\bar{r})$ coordinates. Figure \ref{fig:eta6-pd} shows such a diagram. Note that for a defect free crystal $\bar{r}=r_0$, while for a rigid crystal $\bar{u}=u_0$. The point at which the first order transition turns to second order is the tricritical point designated by open circle in Fig. \ref{fig:eta6-pd}. Figure \ref{fig:Landau-pot} shows a generic sextic Landau potential with coefficients $u<0$ and $v>0$ describing a first order phase transition between the parent-phase and the second-phase, denoted by I and II, respectively.

%
\section{Connection to classical nucleation theory}
\label{sec:cnt}
Classical nucleation theory (CNT) has been used in the past to study the problem of the formation of coherent embryo on or near an edge dislocation \cite{Lyubov_Solovyev_1965,Dollins_1970,Larche_1979}. In this framework, the free energy of the formation can be expressed as \cite{Larche_1979}
\begin{equation}
\label{eqn:cnt-fe}
\Delta F =  -\frac{2\pi}{3}KA R^2 +4\pi\gamma R^2 -\frac{4\pi}{3}\mu_{\mathrm{eff}}R^3,
\end{equation}
\noindent
where $R$ is the radius of the nucleus, $K$ is the bulk modulus, $\gamma$ is the surface tension of the nucleus in the absence of dislocation, and $\mu_\mathrm{eff}$ is the free energy difference per unit volume between the metastable and stable phases. The parameter $A$ was defined earlier, i.e. after Eq. (\ref{eqn:toten1}). Now, $\Delta F$ takes a maximum at $R=R_c$ given by
\begin{eqnarray}
\label{eqn:cnt-fe-crit}
\Delta F_c &=& \frac{4\pi}{3}\Big(\gamma-\frac{KA}{6}\Big)R_c^2,\\
\text{with} \quad R_c &=& \frac{2}{\mu_\mathrm{eff}}\Big(\gamma-\frac{KA}{6}\Big).
\label{eqn:crit-rad}
\end{eqnarray}
\noindent
Note that $A$ has dimension of length and $\gamma$ that of energy per unit area. It is seen that the dislocation simply shifts the surface tension to an effective surface tension $\gamma_\mathrm{eff}=(\gamma-KA/6)$; and so we may write $\Delta F_c=(4\pi/3)\gamma_\mathrm{eff}R_c^2$  and  $R_c=2\gamma_\mathrm{eff}/\mu_{\mathrm{eff}}$.

The surface tension of the nucleus  $\gamma$ is the excess energy stored in the matrix/nucleus interface region per unit area. It can be expressed in terms of the coefficients of the Ginzburg-Landau free energy, cf. Eq. (\ref{eqn:toten1}). Let us for the sake of simplicity ignore the positive stabilizing term $\eta^6$ and put $A=0$. Moreover, consider $r_0<0$ and a planar interface whose normal direction is in the $x$-direction. The Ginzburg-Landau equation becomes
\begin{equation}
\label{eqn:gl-eq2}
g\frac{d^2\eta}{dx^2}+|r_0|\eta-u_1\eta^3=0.
\end{equation}
Then we have a well-known solution for the interface profile
\begin{equation}
\label{eqn:profile}
\eta(x) = \Big(\frac{|r_0|}{u_1}\Big)^{1/2}\tanh\Big(\frac{x}{2\xi}\Big),
\end{equation}
where $\xi=(g/2|r_0|)^{1/2}$ is a characteristic length (the correlation length) for the formation of the new phase. The surface tension is given by
\begin{equation}
\label{eqn:surf-tens-1}
\gamma = \int_{-\infty}^\infty \Big[\frac{g}{2}(\nabla\eta)^2 + \mathcal{V}(\eta)\,\Big]dx.
\end{equation}
 Here $\nabla\eta=d\eta/dx$ and
\begin{equation}
\label{eqn:sympot}
\mathcal{V}(\eta) = \frac{u_1}{4}\Big(\eta^2 - \frac{r_0}{u_1}\Big)^2,
\end{equation}
\noindent
where the free energy density  $-r_0^2/4u_1$ at $x=\pm\infty$ has been deducted from the total energy density. Utilizing now Eq. (\ref{eqn:gl-eq}), after some manipulation, we can write
\begin{equation}
\label{eqn:surf-tens-2}
\gamma = g\int_{-\infty}^\infty (\nabla\eta)^2 dx,
\end{equation}
Next, substituting the interface profile solution Eq. (\ref{eqn:profile}) in (\ref{eqn:surf-tens-2}), then evaluating the integral,  we obtain
\begin{equation}
\label{eqn:surf-tens-3}
\gamma = \frac{2}{3}g  \frac{|r_0|}{u_1}\xi^{-1}=g^{1/2}\,\frac{|2r_0|^{3/2}}{3u_1}.
\end{equation}
Thus, we can express the maximal free energy of formation
\begin{eqnarray}
\label{eqn:cnt-fe-crit}
\Delta F_c &=& \frac{4\pi}{3}\Big[\frac{2}{3}g  \Big(\frac{r_0}{u_1}\Big)\xi^{-1}-\frac{KA}{6}\Big]R_c^2,\\
\text{with} \quad R_c &=& \frac{2}{\mu_\mathrm{eff}}\Big[\frac{2}{3}g  \Big(\frac{r_0}{u_1}\Big)\xi^{-1}-\frac{KA}{6}\Big].
\label{eqn:crit-rad}
\end{eqnarray}
\noindent

On substitution of Eq. (\ref{eqn:crit-rad}) in (\ref{eqn:cnt-fe-crit}) using the formula for $\xi$, the maximal energy of formation can be expressed in the form
\begin{eqnarray}
\label{eqn:cnt-landau-crit}
\Delta F_c &=& \frac{2^{12}\sqrt{2}\pi}{3^4}\frac{(g^3r_0)^{1/2}}{u_1}(1-\varrho)^3\Big(\frac{\mathcal{V}_{\mathrm{max}}}{\mu_\mathrm{eff}}\Big)^2,\\
\text{with} \quad \varrho &=& \frac{1}{4}\frac{u_1}{r_0}\frac{KA}{\sqrt{2g|r_0|}},
\label{eqn:varrho}
\end{eqnarray}
\noindent
and $\mathcal{V}_{\mathrm{max}}=r_0^2/4u_1$ is the maximum value for the Landau potential (\ref{eqn:sympot}). Therefore, the maximal value of the formation energy, in the CNT setting, is proportional to the square of the maximum of $\mathcal{V}(\eta)$. Note also that in the absence of defect $\varrho=0$.

\section{Discussion}
\label{sec:discuss}

Let us first discuss the characteristic length of the system under consideration. We recall the radial part of the bound state solution, Eq. (\ref{eqn:2dse-cef-gs}), namely,
\begin{equation}
\label{eqn:rad-bstate}
\phi(\rho) = \rho^\mu e^{-\rho/\rho_0},
\end{equation}
where $\rho_0=\beta^{-1}$, or more explicitly
\begin{equation}
\label{eqn:rho0}
\rho_0 =\frac{1}{\sqrt{-E_{00}}}=\Big(\frac{2B^2}{r_{00}g}\Big)^{1/2}.
\end{equation}
 To obtain Eq. (\ref{eqn:rho0}), we utilized the relations in (\ref{eqn:2dse-param}) and (\ref{eqn:paradimless2}), with the ground state value of $r_0$, i.e. $r_{00} \propto E_{00}^{-1}$. Scaling $\rho_0$ with $\ell$ to attain dimension of length, we get
\begin{equation}
\label{eqn:xi0}
\xi_0 \equiv \ell\rho_0=\Big(\frac{g}{2|r_{00}|}\Big)^{1/2}.
\end{equation}
The characteristic length $\xi_0$ may be interpreted as the ground state size of the embryo; and since $r_0=\alpha\tau$, we may write $r_{00}=\alpha_0\tau_0$, where $\tau_0=T_0/T_c-1$; thereby $\xi_0=(g/2\alpha_0)^{1/2}|\tau_0|^{-1/2}$, with $T_0$ being the ground-state nucleation temperature. Thus the size of the nucleus at the onset of nucleation is finite \cite{Boulbitch_Toledano_1998}.

A point worth noting is that the Landau mean field theory of phase transition is valid so long as the fluctuations of the order parameter in a volume with linear dimension of order $\xi$ is small compared with the characteristic equilibrium value $\bar{\eta}=(|r_0|/u_1)^{1/2}$. The applicability of the mean field thermodynamics to describe the fluctuations at the onset of nucleation may be checked through the Levanyuk-Ginzburg criterion (see e.g., \cite{Patashinskii_Pokrovskii_1979}), which states
\begin{equation}
\label{eqn:ginzburg-c}
\frac{k_B^2T_c^2u_1^2}{\alpha_0g^3} <<|\tau_0| << 1.
\end{equation}
\noindent
 The ratio on the left-hand side of Eq. (\ref{eqn:ginzburg-c}) is dimensionless ($k_B$ is the Boltzmann constant) and is referred to as the Ginzburg number $Gi$. This number can be expressed in terms of characteristic lengths in the system, i.e. $Gi = (\xi_G/\xi_0)^6$, where
\begin{equation}
\label{eqn:ginzburg-length}
\xi_G = \Big(\frac{k_B^2T_c^2u_1^2}{8\alpha_0|r_{00}|^3}\Big)^{1/6}.
\end{equation}
\noindent
So for the mean field theory to be valid $\xi_G << \xi_0$. We may also express the surface tension in terms of the Ginzburg number in the manner
\begin{equation}
\label{eqn:gamma-gnumber}
\gamma = \frac{\sqrt{2}}{3}k_BT_c\xi_0^{-2}\Big(\frac{|\tau_0|}{Gi}\Big)^{1/2}.
\end{equation}
\noindent
Thus, in this formulation, $\gamma$ is meaningful so long as $|\tau_0|>> Gi$.

Now to make the concepts more tangible, we give a numerical example relevant to precipitation of hydrides in zirconium alloys with a terminal solid solubility for precipitation of about 2 weight parts per million hydrogen at room temperature \cite{Une_Ishimoto_2003}. Typical values for Poisson's ratio $\nu=0.37$, Young's modulus $Y=100$ GPa \cite{Northwood_et_al_1975}, and the magnitude of the Burgers vector $b=0.25$ nm \cite{Bai_et_al_1994} can be used. Furthermore, taking $\kappa/Y=0.1$ m$^4$, $g=10^{-8}$ Jm$^3$, $\alpha_0/T_c=2.2\times 10^5$ JmK$^{-1}$, we find $T_0-T_c \approx 10$ K. With these numerical values, the critical radius for the formation of embryo nucleus, according to Eq. (\ref{eqn:xi0}), turns out as $\xi_0 \approx 50$ nm. Next, the surface tension of the nucleus can be estimated by using Eq. (\ref{eqn:surf-tens-3}). If $u_1 \approx 10^6$ Jm$^5$, then $\gamma \approx 0.3$ Jm$^{-2}$. Finally, with the aforementioned numerical values $Gi=2.6\times 10^{-13}$.

For an application of the model to a particular material, such as precipitation of second phase ZrH$_x$ in Zr matrix, the chemical free energy functional, expressed in terms of the concentration of hydrogen in the matrix, should be coupled to the system of the governing equations. In this context, an additional basic equation, namely the diffusion equation or the Cahn-Hilliard equation, should be solved together with the Ginzburg-Landau equation \cite{Massih_2011a}. Moreover, the effect of appreciable plastic work done by the hydride under stress through volume expansion ($\mathrm{Tr}(\epsilon_{ij}^h)\approx 0.16$) needs to be taken into account. Consequently, even more model parameters will appear in the equations, where their values need to be determined. As noted in section \ref{sec:gov-eqs}, some of the parameters, such as the elasticity constants, the Burgers vector, etc, are directly measurable. Other parameters, such as the coefficients of the Landau expansion may be calculated by \emph{ab initio} type methods. For example, for the precipitation of $\theta^\prime$ phase in defect free Al-Cu alloys such analysis has successfully been carried out \cite{Vaithyanathan_et_al_2004}. For the aforementioned numerical example, the determination of the parameters $\alpha_0$, $g$, $u_1$, and $\kappa$ would be worthwhile.

\section{Concluding remarks}
\label{sec:close}
In this paper we have presented a model for nucleation of second phase at or around dislocation in a crystalline solid. The model employs the Ginzburg-Landau approach for phase transition comprising the sextic term in order parameter ($\eta^6$) in the Landau potential energy. The asymptotically exact solution of linearized time-independent Ginzburg-Landau equation has been found, through which the spatial variation of the order parameter in the vicinity of dislocation has been delineated. Moreover, a generic phase diagram indicating a tricritical behavior near and away from the dislocation is depicted. The relation between the classical nucleation theory and the Ginzburg-Landau approach has been evaluated, for which the critical formation energy of nucleus is related to the maximal of the Landau potential energy. In a numerical example, a number of model parameters is fixed to certain numerical values to obtain plausible results, namely, the shift in the phase transition temperature due to the presence of the dislocation, the critical radius for the formation of embryo nucleus, and the surface tension of the nucleus. The numerical values for the model input constants need to be determined by detailed experiments and/or \emph{ab initio} computations for the system under consideration.

Numerical calculations applicable to the time-dependent Ginzburg-Landau equation with the nonlinear terms will be presented elsewhere. Subsequent steps in our study are an extension of the calculations to a two-component field structural order parameter, the space-time evolution of the order parameter in the vicinity of defects, and the coupling of the composition and the structure order parameters.

\begin{acknowledgments}
 The work was supported by the Knowledge Foundation of Sweden grant number 2008/0503.
\end{acknowledgments}
\appendix

\section{Elimination of elastic field}
\label{sec:appA}
The procedure of eliminating the elastic field appearing in the expressions of the system free energy has been discussed by many authors in the literature, e.g. \cite{Onuki_2002,Sagui_et_al_1994,Leonard_Desai_1998}. Here, we outline a simple procedure applicable to our case. The total free energy functional, Eq. (\ref{eqn:toten}), is expressed in terms of two field variables, the order parameter $\eta$ and the displacement field vector $\mathbf{u}$. In equilibrium, the spatial distribution of the order parameter and the displacement are determined by the Euler-Lagrange equation:
\begin{eqnarray}
\label{eqn:EL1}
\frac{\delta \mathcal{F}}{\delta\eta} &=& 0,\\
\frac{\delta \mathcal{F}}{\delta u_i} &=& Q_i,
\label{eqn:EL2}
\end{eqnarray}
\noindent
where $Q_i$ on denotes the force field due to the presence of an elastic defect in the solid. It has been shown in \cite{Landau_Lifshitz_1970} that this force per unit volume element $dV$ is $\mathbf{Q}/dV=M\boldsymbol{\tau}\times\mathbf{b}\,\delta(\boldsymbol{\xi})$, where $\boldsymbol{\xi}$ is a two-dimensional radius vector perpendicular to $\boldsymbol{\tau}$ with origin at the dislocation line, $\boldsymbol{\tau}$ is a vector parallel to the dislocation line, $\mathbf{b}$ is the Burgers vector, and $\delta(\bullet)$ is the Dirac delta. Thus, by virtue of Eqs. (\ref{eqn:toten})-(\ref{eqn:f-int}), Eqs. (\ref{eqn:EL1}) and (\ref{eqn:EL1}), respectively, yield
\begin{eqnarray}
\label{eqn:EGL-a}
 r_0\eta+u_0\eta^3+v_0\eta^5+2\kappa\eta\nabla\cdot\mathbf{u} &=& g\nabla^2\eta,\\
M\nabla^2 \mathbf{u}+(\Lambda-M)\nabla \nabla \cdot \mathbf{u} + \kappa\nabla \eta^2 &=& M \boldsymbol{\tau}\times\mathbf{b}\delta(\boldsymbol{\xi}),
\label{eqn:mecheq-a}
\end{eqnarray}
where $\Lambda \equiv K+2M(1-1/d)$. For an edge dislocation, $\boldsymbol{\tau}=-\boldsymbol{e}_z$ is constant along the dislocation line, while the Burgers vector is in the $x$ direction $\mathbf{b}=\boldsymbol{e}_x$. Here,  $\boldsymbol{e}_i$ denotes the unit vector along the $i$ axis. Hence the right hand side of Eq. (\ref{eqn:mecheq-a}) for an edge dislocation becomes $-Mb\boldsymbol{e}_y\delta(x)\delta(y)$, which gives Eq. (\ref{eqn:mecheq}). Solving now Eq. (\ref{eqn:mecheq})  the two components of the displacement vector are
\begin{eqnarray}
\label{eqn:disp-x}
 u_x & = & \frac{b}{2\pi}\Big[D\frac{xy}{x^2+y^2}-\arctan\Big(\frac{x}{y}\Big)\Big]-\frac{\kappa}{\Lambda}\nabla_x\nabla^{-2}\eta^2,\\
 u_y & = & \frac{b}{2\pi}\Big[D\frac{y^2}{x^2+y^2}-\frac{M}{2\Lambda}\log\Big(\frac{x^2+y^2}{b^2}\Big)\Big]-\frac{\kappa}{\Lambda}\nabla_y\nabla^{-2}\eta^2,
 \label{eqn:disp-y}
\end{eqnarray}
where $D=(K+M/3)/\Lambda$ and $\nabla^{-2}$ is the inverse Laplacian operator. The dilatation strain $\nabla\cdot \textbf{u}=\nabla_xu_x+\nabla_yu_y$ is (cf. with relations in \cite{Landau_Lifshitz_1970})
\begin{equation}
\label{eqn:dilat}
 \nabla\cdot \mathbf{u} =  -\frac{b}{2\pi}\frac{2M}{\Lambda}\Big(\frac{y}{x^2+y^2}\Big)-\frac{\kappa}{\Lambda}\eta^2.
\end{equation}
\noindent
Inserting Eq. (\ref{eqn:dilat}) into Eq. (\ref{eqn:EGL-a}), we write
\begin{equation}
\label{eqn:EGL-a2}
  g\nabla^2\eta = r_1\eta+u_1\eta^3+v_0\eta^5,
\end{equation}
where renormalized parameters are
\begin{equation}
\label{eqn:renorm-para}
  r_1 = r_0 -\kappa A \frac{\cos\theta}{r},\qquad  u_1 = u_0 -\frac{2\kappa^2}{\Lambda},
\end{equation}
with $A=(2b/\pi)M/\Lambda$, $\theta=\arctan(y/x)$ and $r^2=x^2+y^2$.

\section{Eigenfunctions and eigenvalues of $\hat{H}$: Variational and perturbative method}
\label{sec:appB}
 We use the method proposed by Dubrovskii \cite{Dubrovskii_1997} to solve Eq. (\ref{eqn:paradimless1}). In this method one first writes Eq. (\ref{eqn:lgl-eq}) in the operator form $\hat{H}\psi=E\psi$, where $\hat{H}=\hat{H}_0+\hat{H}_1$, and
\begin{eqnarray}
\label{eqn:H0}
 \hat{H}_0 & = & -\nabla^2 +\frac{\alpha_1-2q\cos\theta}{4\rho^2}-\frac{\alpha_2}{\rho},\\
 \hat{H}_1 & = & \frac{\sin\theta}{\rho}-\frac{\alpha_1-2q\cos\theta}{4\rho^2}+\frac{\alpha_2}{\rho},
 \label{eqn:Hi}
\end{eqnarray}
\noindent
where $q$, $\alpha_1$ and $\alpha_2$ are variational parameters to be determined. Now the variables in the equation $\hat{H}_0\psi=E_0\psi$ are separable. Hence we write $\psi(\rho,\theta)=R(\rho)\Phi(\theta)$, then the following equations are obtained
\begin{eqnarray}
\label{eqn:mathieu}
 \frac{d^2\Phi}{dz^2} + (a-2q \cos{2z})\Phi & = & 0,\\
 \frac{d^2R}{d\rho^2} + \frac{1}{\rho}\frac{dR}{d\rho} + \Big[\Big(E_0+\frac{\alpha_2}{\rho}\Big) - \frac{l^2}{\rho^2}\Big] R & = & 0.
 \label{eqn:2dse}
\end{eqnarray}
\noindent
Here, $\theta=2z+\pi$, $l^2=(a+\alpha_1)/4$ and $a$ is a separation constant. Equation (\ref{eqn:mathieu}) is Mathieu's equation \cite{Abramowitz_Stegun_1964,Whittaker_Watson_1927}; whereas Eq. (\ref{eqn:2dse}) is the 2-dimensional radial Schr\"odinger equation, and its solution depends only on $|l|$ \cite{Zaslow_Zandler_1967}. The periodicity condition of $\theta$ with a period 2$\pi$ is satisfied by the Mathieu functions $ce_m(z,q)$ and $se_m(z,q)$ where $m$ is the order of the functions assuming even integers here; hence solution to Eq. (\ref{eqn:mathieu}) can be expressed as
\begin{equation}
\label{eqn:sol-mathieu}
 \Phi_m  =  \left\{ \begin{array}{ll}
 ce_m(a,z,q), & m=0,2,4,\dots,\\
 se_m(a,z,q), & m=2,4,6,\dots,
 \end{array}\right.
\end{equation}
\noindent
where the notation $ce$ and $se$ comes from cosine-elliptic and sine-elliptic, respectively. One should note that for each $q$, the separation constant $a$ assumes an infinite set of discrete values depending on the index and the parity of the functions; see Appendix \ref{sec:appC}.

The normalized eigenfunction of the radial equation (\ref{eqn:2dse}) is well known, e.g. ref. \cite{Zaslow_Zandler_1967}, and can be expressed in the form
\begin{eqnarray}
\label{eqn:2dse-ef}
 R_{nl}(\rho) &=& \frac{2\beta_n}{(2|l|)!}\Bigg[\frac{(n+|l|-1)!}{(2n-1)(n-|l|-1)!}\Bigg]^{1/2}
(2\beta_n\rho)^{|l|}\exp(-\beta_n\rho)\nonumber \\
&& \times {_1\!F_1}(-n+|l|+1,2|l|+1,2\beta_n \rho),
 \end{eqnarray}

\noindent
where $2\beta_n\equiv \alpha_2/(n-1/2)$, $n=1,2,3,\dots$, and ${_1\!F_1}(a,b,z)$ is the confluent hypergeometric function of the first kind \cite{Abramowitz_Stegun_1964}. Also in this setting, $l=0,1,2,3,\dots,n-1$. The bound state energy levels are
\begin{equation}
\label{eqn:2dse-ev}
 E^{(0)}_n  =  -\frac{\alpha_2^2}{4(n-1/2)^2}=-\beta_n^2.
\end{equation}
\noindent

Thus the composite eigenfunction for the system under consideration is
\begin{equation}
\label{eqn:2dse-cef}
 \psi_{mnl}  =  \Phi_m(z,q) R_{nl}(\rho).
\end{equation}
\noindent

In general the eigenfunctions and the eigenvalues of the Hamiltonian $\hat{H}_0$, see Eq. (\ref{eqn:H0}), can be classified by three characteristic numbers $\{n,p,m\}$, where $n=0,1,2,\dots$ is the radial solution index, $p=0,1$ the parity index, determining the inversion symmetry with respect to variable $z$, and $m$ is the index of the Mathieu function, taking up only even integers, viz., $m=0,2,4,\dots$ for $p=0$ and $m=2,4,6,\dots$ for $p=1$. The eigenfunctions of $\hat{H}_0$ can be expressed in the form
\begin{eqnarray}
\label{eqn:2dse-gen-ef}
 \psi_{npm}  &=&  \frac{A_{npm}}{\sqrt{\pi}} \rho^{\mu} \exp(-\beta_{npm}\rho){_1\!F_1}(-n,2\mu+1,2\beta_{npm}\rho)\Phi_{npm},\\
\label{eqn:2dse-gen-par1}
 \text{where} \quad \Phi_{npm} &=& \delta_{p,0}ce_m+\delta_{p,1}se_m,\\
 \beta_{npm} &=& \frac{\alpha_2}{2\mu+2n+1},\\
 \mu &=& \frac{1}{2}(\textbf{a}+\alpha_1)^{1/2},\\
 \textbf{a} &=& \delta_{p,0}a_m+\delta_{p,1}b_m,\\
 A_{npm} &=& \frac{\sqrt{2}(2\beta_{npm})^{1+\mu}}{\Gamma(2\mu+1)}\sqrt{\frac{\Gamma(2\mu+n+1)}{n!(2\mu+n+1)}},\\
 \label{eqn:2dse-gen-ev}
 \text{and} \quad E_{npm}^{(0)} &=& -\beta_{npm}^2,
\end{eqnarray}
\noindent
where $E_{npm}^{(0)}$ define the energy eigenvalues for the Hamiltonian \cite{Nabutovskii_Shapiro_1978,Dubrovskii_1997}. Considering the ground state  $ \psi_{000}$ with $\beta=\beta_{000}=\alpha_2/(2\mu+1)$ and $\mu=(a_0+\alpha_1)^{1/2}/2$, and comparing with the relations in Eq. (\ref{eqn:2dse-param}), the relations in Eq. (\ref{eqn:2dse-gen-par1-gs}) follow.

Next, using standard perturbation technique, the first (non-zero) shift is calculated through (\ref{eqn:2dse-gen-ev})
\begin{equation}
\label{eqn:2dse-e02}
 E_{000}^{(2)}  =  E_{000}^{(0)}(q_0)+\frac{\big(\langle 000|\hat{H}_1|002\rangle\big)^2 }{E_{000}^{(0)}(q_0)-E_{002}^{(0)}(q_2)}
 \end{equation}
 \noindent
with
\begin{equation}
\label{eqn:2dse-e2v}
 \langle 000|\hat{H}|002\rangle \equiv \int_0^\infty \rho d\rho \int_{-\pi/2}^{\pi/2} \psi_{000}(q_0,\rho,z) \hat{H}\psi_{002}(q_2,\rho,z)dz.
 \end{equation}
 \noindent
Here, $q_0$ and $q_2$ are the the minima of $E_{000}^{(0)}(q)$ and  $E_{002}^{(0)}(q)$, respectively, calculated through Eq. (\ref{eqn:2dse-gen-ev}) with numerical values listed below

\begin{center}
\begin{ruledtabular}
\begin{tabular}{l|cccc}
 State	& $q_{\mathrm{min}}$ & $\alpha_1$ & $\alpha_2$ & $E_{npm}^{(0)}(q_{\mathrm{min}})$ \\
 \hline
$|000\rangle$ & 3.91792 & 5.80765 & 0.741165 & -0.105288\\
$|002\rangle$ & 2.76301 & 3.77263 & 0.682702 & -0.0277153\\
\end{tabular}
\end{ruledtabular}
\end{center}
Hence, Eq. (\ref{eqn:2dse-e2v}) gives $\langle 000|\hat{H}|002\rangle = 0.0221$ and Eq. (\ref{eqn:2dse-e02}) yields $E_{000}^{(2)}=-0.112$. Our calculated numerical values agree well with those reported by Dubrovskii \cite{Dubrovskii_1997} for the ground state, but differ somewhat for the first excited state.

\section{Angular Solutions}
\label{sec:appC}

The angular solutions $\Phi_m(z,q)$, Eq. (\ref{eqn:sol-mathieu}), expressed in terms of the Mathieu functions $ce_m(a,q,z)$ and $se_m(a,z,q)$ have subtle properties. The value of $a$ in Eq. (\ref{eqn:mathieu}) form the eigenvalues of the equation; for $ce_m$ they are commonly denoted by $a_m(q)$, whereas for $se_m$ by $b_m(q)$. Hence we write: $a\Rightarrow a(p,m)=\delta_{p,0}a_m+\delta_{p,1}b_m$ with $p=0,1$ and $\delta_{p,q}$ denoting the Kronecker $\delta$. The index $m$ takes up even integers starting from zero for $p=0$ and from 2 for $p=1$. The Mathieu functions form a complete set of functions in the interval $0<z<2\pi$, namely
\begin{equation}
\label{eqn:orthonorm}
 \int_0^{2\pi}ce_m\,ce_n dz  = \int_0^{2\pi}se_m\,se_n dz=\pi\delta_{m,n}
\end{equation}

Finally, since the Mathieu functions $ce_m$ and $se_m$ are periodic, they can be expanded by Fourier series
\begin{eqnarray}
\label{eqn:math-fourier}
 ce_{2r}(a,q,z)  &=& \sum_{k=0}^{\infty}\mathcal{A}_{2k}(q)\cos(2kz)\\
 se_{2r+2}(a,q,z)  &=& \sum_{k=0}^{\infty}\mathcal{B}_{2k+2}(q)\sin[(2k+2)z]
\end{eqnarray}
where $r=0,1,2,\dots$. The recursion relations between the coefficients are obtained by substituting these series into Eq. (\ref{eqn:mathieu}). For example, for $ce_{2r}$, we obtain
\begin{eqnarray}
\label{eqn:recursive}
  q \mathcal{A}_{2}  &=& a\mathcal{A}_{0}\\
  q \mathcal{A}_{4}  &=& (a-4)\mathcal{A}_{2}-2q \mathcal{A}_{0}\\
  q \mathcal{A}_{2k+2} &=& (a-4k^2) \mathcal{A}_{2k}-q \mathcal{A}_{2k-2}
\end{eqnarray}
with $k \ge 2$. More on Mathieu functions is found in \cite{Abramowitz_Stegun_1964,Frenkel_Portugal_2001}.

\newpage
\bibliography{micro}

\end{document}